\documentclass[a4paper,12pt]{article}
\usepackage{graphicx, graphics}
\usepackage{latexsym}
\usepackage{epsfig}
\usepackage{cite}
\usepackage{epstopdf}
\usepackage{amssymb}
\usepackage{epsfig}
\usepackage{amssymb}
\author{Parviz Goodarzi \footnote{E-mail: parviz.goodarzi@abru.ac.ir}
\\ {\small\small Department of Physics, Faculty of Basic Sciences,}
\\ {\small Ayatollah Boroujerdi University, Boroujerd, Iran}}
\title {Intermediate inflation in generalized non-minimal derivative coupling model}
\begin{document}
\maketitle
\begin{abstract}
In this work, we consider intermediate inflation in context of the Generalized Non-Minimal Derivative Coupling (GNMDC) model.
In this model, inflation is driven by a canonical scalar field that is coupled not only to gravity but also to the derivative of the scalar field.
The GNMDC model introduces new dynamics and features during the inflationary epoch.
We find inflationary solutions with a power law scalar field for the power law coupling function.
Additionally, we determine the inflaton potential that generates intermediate expansion of the scale factor.
We also discuss the background equations in the high friction limit and derive constraints on the parameters of our model.
Furthermore, we investigate the cosmological perturbations in the slow roll approximation within the GNMDC model,
We calculate the scalar and tensor spectral index and the tensor-to-scalar ratio during the intermediate inflation.
We compare the results of this model with the observational data that can be used to test the model using the cosmic microwave background radiation data.
Overall, we establish conditions for the inflaton potential that ensure the continuation of accelerated expansion during the slow roll inflation.
We numerically analyze the power spectrum and spectral index for scalar and tensor modes in intermediate inflation in the high friction limit.
Moreover, we use the Planck 2018 data, to obtain constraints on the parameters of the model.
We demonstrate that intermediate inflation in the GNMDC model is successful in evaluation and explanation of the background and perturbational quantities using observational data.
\end{abstract}

\section{Introduction}

One of the significant achievements of the inflationary paradigm is its ability to explain the observed inhomogeneity of the cosmic microwave background radiation and the large-scale structure of the universe \cite{Mukhanov,Olive}. It provides an explanation for the flatness problem, the horizon problem and the magnetic monopole problem, which were not adequately addressed by the original Big Bang theory \cite{guth, Linde1}.
According to the inflationary models, the universe experienced a phase of extremely rapid expansion driven by a hypothetical scalar field called the inflaton \cite{Linde2,Linde3,kolb}. This expansion caused the universe to grow exponentially, stretching out any pre-existing irregularities or fluctuations, and making the universe appear smooth and homogeneous on large scales \cite{Lyth,Liddle,Liddle1}.
Overall, cosmological inflation provides a compelling framework for understanding the early universe's dynamics, its large-scale properties and the origin of structure in the Universe \cite{Odintsov1}.
Many experiments and observations such as those conducted by the Planck satellite and ongoing research on cosmic microwave background radiation, gravitational waves and the large-scale structure of the universe, seek to provide further insights into the validity of specific parameters of inflationary models \cite{Bardeen,Martin}.
There are several models of inflation in which, a canonical scalar field is minimally or non-minimally coupled to gravity, or the inflaton field interacts fundamentally with other fields\cite{Odintsov2,Odintsov3,Mukhanov1,Capozziello,Peebles}.
There are two ways to consider inflationary solutions: first, finding the inflaton potential by assuming that the scale factor is known, or by obtaining the suitable scale factor for a specific potential. In some inflationary models, exact solutions can be found within the framework of general relativity when the evolution of the scale factor follows an exponential growth pattern due to a constant potential, similar to the quasi-de Sitter expansion of the universe \cite{Stephani}. On the other hand, an exponential potential leads to power law inflation, in which the scale factor grows $a(t)\propto t^p$ with $p>1$ \cite{Lucchin}.

Another type of exact solution is intermediate inflation, in which the evolution of the scale factor is expressed
by the relation $a(t)=a_0 \exp(At^{\lambda})$, where $A$ and $0<\lambda<1$ are two constants \cite{Barrow, Barrow1}.
The evolution of the scale factor in intermediate inflation is slower than the de Sitter expansion of the universe, but faster than power law inflation \cite{Herrera}.
Originally, this inflationary model studied in context of exact solution to the background equations but from the observational point of view, intermediate inflation was noticed in the slow roll approximation \cite{Herrera,Herrera1,Herrera2}.

Intermediate inflation is in good agreement with the slow roll approximation in the simple inflation model with a canonical scalar field.
Therefore, many authors have investigated intermediate inflation in different models, such as warm inflation, brane inflation, G-inflation, and non-canonical scalar field \cite{Herrera3,Herrera4,Herrera5,Rezazadeh,Teimoori}.
In reference \cite{Rashidi} intermediate inflation in the tachyon model with constant sound speed has been examined.
Also, \cite{Herrera} intermediate inflation in a generalized non-minimal coupling to the scalar curvature is investigated.
The observational viability of the intermediate inflation within the context of the Galileon scenario has been considered in \cite{Herrera2,Teimoori}.
The warm-intermediate inflationary model is examined in the presence of the Galileon coupling \cite{Herrera4}.
General conditions required for intermediate inflation are discussed from the background and cosmological perturbations in the slow-roll regime \cite{Herrera3}.

The aim of this paper is to examine intermediate inflation in the generalized non-minimal derivative coupling model.
Generalized non-minimal kinetic coupling $f(\varphi)G^{\mu \nu}\partial_\mu\varphi \partial_{\nu} \varphi$ is an interesting operator of Horndeski’s scalar-tensor theory which primordially introduced in \cite{Horndeski,Charmousis}.

An emphasize feature of Horndeski’s scalar-tensor theory is that in spite of higher order terms in the Lagrangian, their corresponding field equations are second order and do not produce ghost instability. Furthermore, the Horndeski theory include a wide rang of gravitational theories, such as non-canonical scalar field models, generalized G-inflation, Brans-Dicke theory \cite{Nakayama,Kobayashi,Deffayet,Deffayet1,Brans}.
The more general Horndeski’s scalar-tensor theory which lagrangian includes quadratic terms, is the only theory in which the anisotropies are damped at early times \cite{Sushkov1,Sushkov2}.

Non-minimal derivative coupling model is the simplest case of this model with the constant coupling function $f(\varphi)=1/M^2$ where used in references \cite{Germani1,Germani2} to explain Higgs inflation.
Observational tests, reheating period, reheating temperature, gravitational baryogenesis, warm inflation, exact cosmological solutions, quintessence and phantom cosmology, have been investigated in the non-minimal derivative coupling model \cite{Tsujikawa,Sadjadi,Sadjadi1,Parviz,Sushkov3,Sushkov4,Sushkov5,Sushkov6}.

However, finding exact solutions for intermediate inflation in this model is not possible due to the presence of the constant coupling function.
In the GNMDC model, the coupling function provides the possibility of finding the exact solutions for intermediate inflation.
In GNMDC model the inflaton field evolves more slowly relative to the case of standard canonical inflation due to a gravitationally enhanced friction which its capacity to explain inflation for any steep inflaton potential \cite{Germani3}.
 Moreover this coupling is also safe of quantum corrections and unitary violation problem, without introducing new degrees of freedom \cite{Germani3}.
The inflationary prediction of GNMDC and primordial black hole production where fully investigated in \cite{Dalianis,Chengjie}.
They examined the inflationary phenomenologically with the Higgs potential and exponential potential.
An emphasize feature of the GNMDC with the Einstein tensor is that the mechanism of the gravitationally enhanced friction during inflation, by which wide range of potentials with theoretically parameters where they are acceptable in terms of observations, can drive cosmic acceleration \cite{Germani3, Tsujikawa}.
Consequently, this motivates us to investigate intermediate inflation in context of GNMDC in light of the Planck 2018 results.
In this paper, we will examine intermediate inflation in the presence of the power law potential and power law coupling function within the context of the generalized non-minimal derivative coupling model.
The effect of "gravitationally enhanced friction" on the evolution of the scalar field during inflation and its impact on the observational parameters of the early Universe will be examined .
We will consider the spectral index, power spectrum, the number of e-folds and the tensor to scalar ratio in the intermediate inflation in framework of GNMDC analytically and numerically.

This paper is organized as follows.
In section 2 we review the generalized non-minimal derivative coupling in FLRW geometry, and we obtain the basic equations of motion for the scalar field.
In section 3 we derive intermediate inflation in the slow roll approximation for equations of motion, with the power-law coupling function.
In section 4 we consider cosmological perturbations by using slow roll approximation within GNMDC model.
We obtain power spectrum and spectral index for scalar and tensor modes for intermediate inflation.
In section 5 we numerically investigate the evaluation of the model's parameters for different values of $\lambda$, in the high friction regime.
There are interesting solutions for different values of the parameters and we compare the numerical results with the observational data.
Finally, conclusion and brief discussions are given in the final section.

We use units $\hbar=c=1$ though the paper.

\section{Generalized non-minimal derivative coupling}

In this section, we will investigate the evolution of the scalar field by the mechanism of gravitationally enhanced friction in the context of non minimal derivative coupling between gravity and the inflaton field.
An action of the GNMDC theory in the Jordan frame is given by \cite{Chengjie,Dalianis}

\begin{equation}\label{1}
S=\int \Big({M_P^2\over 2}R-{1\over 2}(g^{\mu \nu}-f(\varphi)G^{\mu \nu})\partial_\mu
\varphi \partial_{\nu} \varphi- V(\varphi)\Big) \sqrt{-g}d^4x,
\end{equation}

where $G^{\mu \nu}=R^{\mu \nu}-{1\over 2}Rg^{\mu \nu}$ is Einstein tensor, $R$ is Ricci scalar, $f(\varphi)$ is the coupling function, $V(\varphi)$ is the inflaton potential, and $M_P=2.4\times 10^{18}Gev$ is the reduced Planck mass.
We can obtain the field equation and the energy momentum tensor by variation of the action (\ref{1}) with respect to the metric $g_{\mu\nu}$, as
\begin{eqnarray}
\label{2}
G_{\mu\nu}&=&{1\over M_p^2}T_{\mu\nu}, \\
\label{3}
T_{\mu\nu}&=&T^{(0)}_{\mu\nu}-f(\varphi)T^{(1)}_{\mu\nu}-{1\over2}f'(\varphi)T^{(2)}_{\mu\nu},
\end{eqnarray}
where $f'(\varphi)={df/d\varphi}$.
The energy momentum tensor for minimal and non-minimal coupling counterparts of scalar field as follows
\begin{eqnarray}\label{4}
T^{(0)}_{\mu\nu}&=&\nabla_{\mu}\varphi\nabla_{\nu}\varphi-{1\over2}g_{\mu\nu}{(\nabla\varphi)}^2-g_{\mu\nu}V(\varphi),\\\nonumber
T^{(1)}_{\mu\nu}&=&-G_{\mu\nu}{(\nabla\varphi)}^2-R\nabla_{\mu}\varphi\nabla_{\nu}\varphi\\\nonumber
&&+2\Big(R^{\alpha}_{\mu}\nabla_{\alpha}\varphi\nabla_{\nu}\varphi+R^{\alpha}_{\nu}\nabla_{\alpha}\varphi\nabla_{\mu}\varphi\\\nonumber
&&+R_{\mu\alpha\nu\beta}\nabla^{\alpha}\varphi\nabla^{\beta}\varphi
+\nabla_{\mu}\nabla^{\alpha}\varphi\nabla_{\nu}\nabla_{\alpha}\varphi-\nabla_{\mu}\nabla_{\nu}\varphi\Box \varphi\Big)\\\nonumber
&&+g_{\mu\nu}\Big(\nabla^{\alpha}\nabla^{\beta}\varphi\nabla_{\alpha}\nabla_{\beta}\varphi+
{(\Box\varphi)}^2-R^{\alpha\beta}\nabla_{\alpha}\varphi\nabla_{\beta}\varphi\Big),\\\nonumber
T^{(2)}_{\mu\nu}&=&g_{\mu\nu}\Big(\nabla_{\alpha}\varphi\nabla^{\alpha}\varphi\nabla^2\varphi
-\nabla_{\alpha}\varphi\nabla_{\beta}\varphi\nabla^{\alpha}\nabla^{\beta}\varphi\Big)\\\nonumber
&&+\nabla^{\alpha}\varphi\nabla_{\mu}\varphi\nabla_\nu\nabla_\alpha\varphi+\nabla^{\alpha}\varphi\nu_{\alpha}\varphi\nabla_\mu\nabla_\alpha\varphi\\\nonumber
&&-\nabla_{\alpha}\varphi\nabla^{\alpha}\varphi\nabla_{\nu}\nabla_{\mu}\varphi-\nabla_{\mu}\varphi\nabla_{\nu}\varphi\Box\varphi.
\end{eqnarray}
Also, we can obtain equation of motion by variation of the action (\ref{1}) with respect to scalar field
\begin{eqnarray}\label{5}
-g\Big[\Big(\partial_\mu g^{\mu\nu}-f(\varphi)\partial_\mu G^{\mu\nu}\Big)\partial_\nu\varphi&+& \Big(g^{\mu\nu}-f(\varphi)G^{\mu\nu}\Big)\partial_\mu\partial_\nu\varphi \\ \nonumber
-{1\over2}f'(\varphi)G^{\mu\nu}\partial_\mu\varphi\partial_\nu\varphi-V'(\varphi)\Big]&-&
{1\over2}\Big(g^{\mu\nu}-f(\varphi)G^{\mu\nu}\Big)\partial_\nu\varphi\partial_\mu g=0.
\end{eqnarray}

We have seen, that for the constant coupling function $f(\varphi)=1/M^2$, the third term $T^{(2)}_{\mu\nu}$, vanishes and the field equations of the usual non-minimal derivative coupling are obtained \cite{Germani1}.
In the absence of terms containing more than two time derivatives, this theory does not produce additional degrees of freedom.
For a homogeneous scalar field $\varphi=\varphi(t)$, in the FLRW metric we can calculate the Friedman equations from relation (\ref{2}) as
\begin{eqnarray}
\label{6}
3M_p^2H^2&=&\Big(1+9H^2f(\varphi)\Big){\dot{\varphi}^2\over 2}+V(\varphi),\\
\label{7}
M_p^2\dot{H}&=&-{\dot{\varphi}^2\over2}+f(\varphi)\Big[H\dot{\varphi}\ddot{\varphi}-(3H^2-2\dot{H}){\dot{\varphi}^2\over2}
\Big]+{1\over2}f^\prime(\varphi)H\dot{\varphi}^3,
\end{eqnarray}
where an overdot represents differentiation with respect to cosmic time and $H(t)=\dot{a}/a$ is the Hubble parameter.
The effective energy density and pressure for the homogeneous scalar field can be expressed as
\begin{eqnarray}
\label{8}
\rho_\varphi&=&\Big(1+9H^2f(\varphi)\Big){\dot{\varphi}^2\over 2}+V(\varphi),\\
\label{9}
P_\varphi&=&\Big(1-f(\varphi)(3H^2+2\dot{H})\Big){\dot{\varphi}^2\over2}-V(\varphi)\\ \nonumber
&-&2Hf(\varphi)\dot{\varphi}\ddot{\varphi}-f^\prime(\varphi)H\dot{\varphi}^3.
\end{eqnarray}

The scalar field equation of motion from equation (\ref{5}) in FLRW geometry becomes
\begin{eqnarray}\label{10}
\Big(1+3f(\varphi)H^2\Big)\ddot{\varphi}&+&3H\Big(1+3f(\varphi)H^2+2f(\varphi)\dot{H}\Big)\dot{\varphi}\\ \nonumber
&+&{3\over2}f^\prime(\varphi){\dot{\varphi}}^2H^2+V'(\varphi)=0.
\end{eqnarray}
We have seen, that the equation of motion for standard minimal coupling is obtained when $f(\varphi)=0$ in the spatial case.

\section{Intermediate inflation in GNMDC}

In the following we would like to consider, background equations of the GNMDC model, in the slow roll approximation.
We define the slow roll parameters as follows:
\begin{equation}\label{11}
\epsilon\equiv-{\dot{H}\over H^2},\;\;\;  \delta\equiv{\ddot{\varphi}\over H\dot{\varphi}},\;\;\;  \eta\equiv{{\dot{\varphi}}^2\over2 M_p^2H^2}.
\end{equation}
If we define $\mathcal{A} \equiv f(\varphi)H\dot{\varphi}^2$, then another slow roll parameter becomes
\begin{equation}\label{12}
\delta_{\mathcal{A}}\equiv {\dot{\mathcal{A}}\over 3H\mathcal{A}}.
\end{equation}

Different types of $f(\varphi)$ have been considered in the literature \cite{Chengjie,Dalianis}.
Especially, we have used the case where the coupling function is a power law type, in the form of
\begin{equation}\label{13}
f(\varphi)={\varphi^{\alpha-1}\over M^{\alpha+1}}.
\end{equation}

The slow roll regime can be characterized by the set of slow roll conditions
\begin{equation}\label{14}
\{\epsilon,  \delta,  \eta,  \delta_\mathcal{A}\}\ll1.
\end{equation}

In the first step, we can rewrite equation (\ref{7}) in the form of
\begin{equation}\label{15}
2M_p^2\dot{H}=-\dot{\varphi}^2+{d\over dt}\big[f(\varphi)H\dot{\varphi}^2\big]-3H\big[f(\varphi)H\dot{\varphi}^2\big],
\end{equation}
by applying the slow roll approximation (\ref{14}), in equation (\ref{15}), we have
\begin{equation}\label{16}
2M_p^2\dot{H}\approx-\dot{\varphi}^2-3H\big[f(\varphi)H\dot{\varphi}^2\big].
\end{equation}

Additionally, intermediate inflation is specified by the following scale factor \cite{Barrow,Barrow1}
\begin{equation}\label{17}
a(t)=a_0 \exp(At^\lambda),
\end{equation}
where $A>0$ and $0<\lambda<1$ are two constant. Therefore the Hubble parameters becomes
\begin{equation}\label{18}
H\equiv{\dot{a}\over a}=A\lambda t^{(\lambda-1)}.
\end{equation}
Throughout this paper, we assume that the scalar field is a power law function of the cosmic time
\begin{equation}\label{19}
\varphi(t)=c t^{m},
\end{equation}
where the parameters $m$ and $c$ both are constant. By substituting of relations (\ref{19}) and (\ref{18}) into the equation (\ref{6}) we have
\begin{eqnarray}\label{20}
2M_p^2A\lambda(\lambda-1)t^{(\lambda-2)}&\approx&-(cm)^2t^{(2m-2)} \\ \nonumber
&&-3{c^{\alpha-1}\over M^{\alpha+1}}(A\lambda)^{2}(cm)^2t^{(m\alpha+2\lambda+m-4)}.
\end{eqnarray}
The power of $t$ needs to be equal on both sides of this equation, we obtain
\begin{equation}\label{21}
\lambda-2=2m-2=m\alpha+2\lambda+m-4,
\end{equation}
and the coefficient of $t$ needs to be equal on both sides of this equation, we get
\begin{equation}\label{22}
2M_p^2A\lambda(\lambda-1)\approx-(cm)^2-3{c^{\alpha-1}\over M^{\alpha+1}}(A\lambda)^{2}(cm)^2.
\end{equation}
Therefore, from relation (\ref{21}), we can write the parameters $m$ and $\alpha$ in terms of $\lambda$ as
\begin{equation}\label{23}
m={\lambda\over2}, \;\;\; \alpha={4-3\lambda\over \lambda}.
\end{equation}

As we have seen, the constraint $0<\lambda<1$ indicates that, in order to have inflation, we need $0<m<0.5$ and $\alpha>1$.
Now, by substituting of $m$ and $\alpha$ from relation (\ref{23}) into equation (\ref{22}), we can rewrite this equation follow as
\begin{equation}\label{24}
8M_p^2A(\lambda-1)\approx-\Bigg(1+3(A\lambda)^{2}{c^{4\big({1-\lambda\over\lambda}\big)}\over M^{2\big({2-\lambda\over\lambda}\big)}}\Bigg)c^2\lambda.
\end{equation}

To calculate the effective inflaton potential $V(\varphi)$ in terms of the scalar field, we utilize the Hamiltonian constraint (\ref{6}) in the slow roll approximation as
\begin{eqnarray}\label{25}
V(\varphi)&\approx& 3M_p^2(A\lambda)^2\Big({\varphi\over c}\Big)^{4\big({\lambda-1\over\lambda}\big)} \\\nonumber
&-&{(c\lambda)^2\over8}\Bigg(1+9(A\lambda)^{2}{c^{4\big({1-\lambda\over\lambda}\big)}\over M^{2\big({2-\lambda\over\lambda}\big)}}\Bigg)\Big({\varphi\over c}\Big)^{2\big({\lambda-2\over\lambda}\big)}.
\end{eqnarray}

Another essential quantity to study cosmological inflation is the number of e-folds $\mathcal{N}$, defined as
\begin{eqnarray}\label{26}
\mathcal{N}\equiv\int_{t_{0}}^{t_{end}}Hdt&=&A({t_{end}}^\lambda-{t_{0}}^\lambda) \\ \nonumber
&=&{A\over c^2}({\varphi_{end}}^2-{\varphi_0}^2),
\end{eqnarray}
where $\varphi_{end}=\varphi(t_{end})$ and $\varphi_{0}=\varphi(t_0)$ are the values of the scalar
field at the end and the beginning of the slow roll inflation, respectively.
Now, one can calculate $\mathcal{A}$ from relations (\ref{13}), (\ref{18}) and (\ref{19}) as a function of cosmic time

\begin{equation}\label{26.1}
\mathcal{A}={m^2A\lambda\over t}\Big({c\over M}\Big)^{\alpha+1}={2M_p^2(1-\lambda)\over3t}.
\end{equation}

From relation (\ref{11}) and (\ref{12}) we can obtain the slow roll parameters as a function of cosmic time or scalar field as
\begin{eqnarray}
\label{27}
\epsilon&=&-{(\lambda-1)\over A\lambda t^\lambda}={c^2(1-\lambda)\over A\lambda}{1\over \varphi^2}, \\
\label{27.1}
\eta&=&{c^2\over 8M_p^2A^2 t^\lambda}={c^4\over 8M_p^2A^2}{1\over \varphi^2}, \\
\label{27.2}
\delta&=&-{(2-\lambda)\over 2A\lambda t^\lambda}={c^2(\lambda-2)\over 2A\lambda}{1\over \varphi^2}, \\
\label{28}
\delta_{\mathcal{A}}&=&-{1\over 3A\lambda t^\lambda}=-{c^2\over 3A\lambda}{1\over \varphi^2}.
\end{eqnarray}

As you can see, in the intermediate inflation, contrary to the power law inflation where the slow roll parameters increase during inflation, the slow roll parameters $\epsilon$ decreases at all times.
Therefore, determining the time of the beginning and the end of inflation is a significant issue.
In view of the fact that the slow roll parameter $\epsilon$ is a decreasing function, we choose $\epsilon\approx 1$ at the beginning of inflation so that after this time the slow roll parameter becomes smaller than unity $(\epsilon<1)$, in other words the expansion of the universe becomes accelerated.
Therefore, the beginning of inflation and the scalar field at beginning of inflation becomes
\begin{equation}\label{29}
{t_0}^\lambda\approx \Big({{1-\lambda}\over \lambda}\Big){1\over A}, \;\;\; {\varphi_{0}}^2\approx \Big({{1-\lambda}\over \lambda}\Big){c^2\over A}.
\end{equation}

In order to address the issues of flatness, horizon, and other problems in the early Universe, inflation must continue until the number of e-folds reaches $\mathcal{N}>60$.
Then, the scalar field at the end of inflation is given in terms of the initial values of the scalar field and the number of e-folds in the form of

\begin{equation}\label{30}
{t_{end}}^\lambda\approx \Big(\mathcal{N}+{1-\lambda\over \lambda}\Big){1\over A}, \;\;\; {\varphi_{end}}^2\approx\Big(\mathcal{N}+{{1-\lambda}\over \lambda}\Big){c^2\over A}.
\end{equation}
We have seen $\varphi_{end}>\varphi_{0}$.
The high friction condition in GNMDC model is
\begin{equation}\label{31}
f(\varphi)H^2\gg 1.
\end{equation}
By inserting relations (\ref{18}) and (\ref{19}) into condition (\ref{31}) one can obtain the high friction condition, in the following form
\begin{equation}\label{32}
M^{(\alpha+1)}\ll(A\lambda)^2 c^{(\alpha-1)}.
\end{equation}

Now, in this regime, we can calculate $c$ from relation (\ref{22}) as
\begin{equation}\label{33}
c\approx\Bigg({8(1-\lambda)\over3A\lambda^3}M_p^2\Bigg)^{\lambda\over2(2-\lambda)}M.
\end{equation}
Therefore, by utilizing equation (\ref{33}), we can express the high friction limit (\ref{32}) in two different following forms
\begin{eqnarray}\label{33.1}
 c&\ll&\sqrt{{8(1-\lambda)A\over3\lambda}}M_p, \\
\label{33.2}
 M&\ll&\Bigg({8(1-\lambda)A\over3\lambda}M_p^2\Bigg)^{1-\lambda\over2-\lambda}(A\lambda)^{\lambda\over(2-\lambda)}.
\end{eqnarray}

Furthermore, in this regime, the potential energy of the inflaton (\ref{25}) becomes
\begin{equation}\label{34}
V(\varphi)=3M_p^2 A\lambda\Big({\varphi\over c}\Big)^{2(\lambda-2)\over\lambda}\Big[(\lambda-1)+A\lambda\Big({\varphi\over c}\Big)^2\Big].
\end{equation}

It is worth mentioning that the evaluation of this potential at the beginning of inflation (\ref{29}) is zero $V(\varphi_0)=0$.
Subsequently, we can obtain the potential at the end of inflation as
\begin{equation}\label{34.1}
V(\varphi_{end})=3M_p^2A^{2\over\lambda}\lambda^{(\lambda+2)\over\lambda}(\mathcal{N}\lambda-\lambda+1)^{(\lambda-2)\over\lambda}\mathcal{N}.
\end{equation}

Furthermore, the highest potential value is located at
\begin{equation}\label{34.2}
\varphi_{max}=c\sqrt{2-\lambda\over 2A\lambda},
\end{equation}
thus, the maximum value of the potential becomes
\begin{equation}\label{34.3}
V(\varphi_{max})={3\over 2}M_p^2 A\lambda^2\Big({2-\lambda\over 2A\lambda}\Big)^{(\lambda-2)\over\lambda}.
\end{equation}

\begin{figure}
\begin{center}
  \scalebox{0.49}{\includegraphics{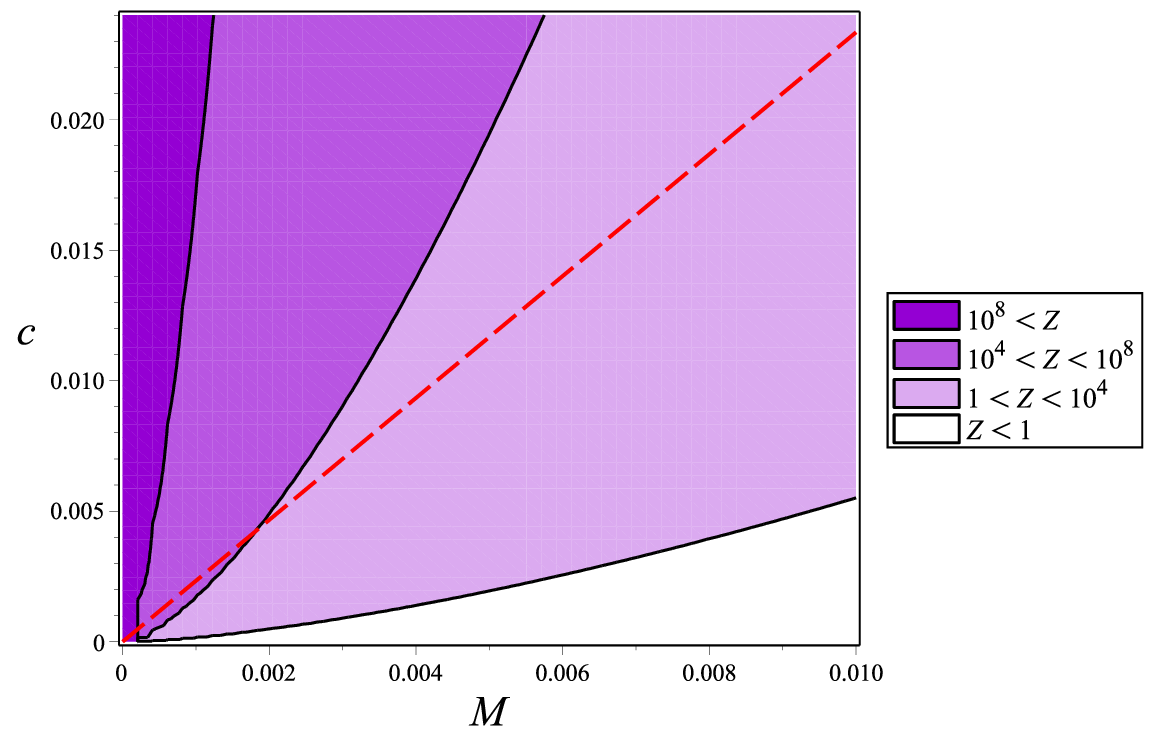}}
  \scalebox{0.5}{\includegraphics{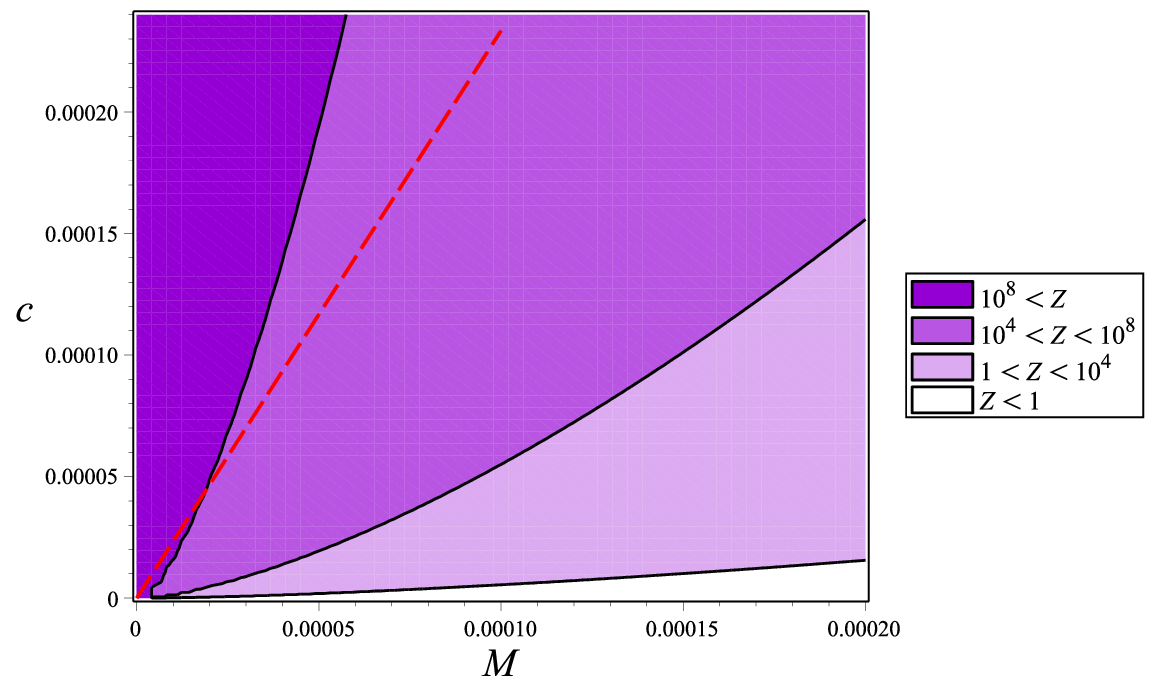}}
  \scalebox{0.5}{\includegraphics{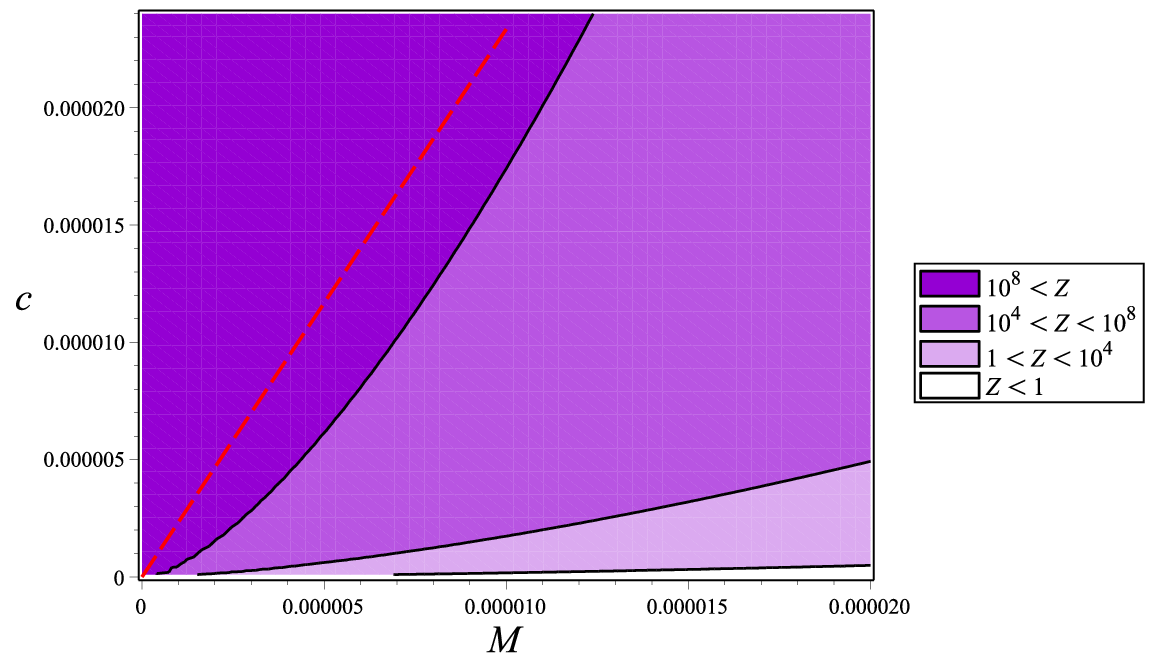}}
  \end{center}
   \caption{\footnotesize The admissible regions for the values of the parameters $c$ and $M$ are restricted to different shades of violet color with $\lambda=0.5$, due to the high friction constraint $(Z\gg1)$.
   The regions with bright colors, represent low values of $Z$, while the regions with dark colors represent high values of $Z$.
   The acceptable values for parameters $c$ and $M$ from relation (\ref{33}) are shown with a red dashed line.
   The best values for parameters $c$ and $M$ are indicated by the red dashed line, which is located within the shaded region.}
\label{fig1}
\end{figure}

\section{Cosmological perturbation in GNMDC}

\begin{figure}[t]
\begin{center}
  \scalebox{0.5}{\includegraphics{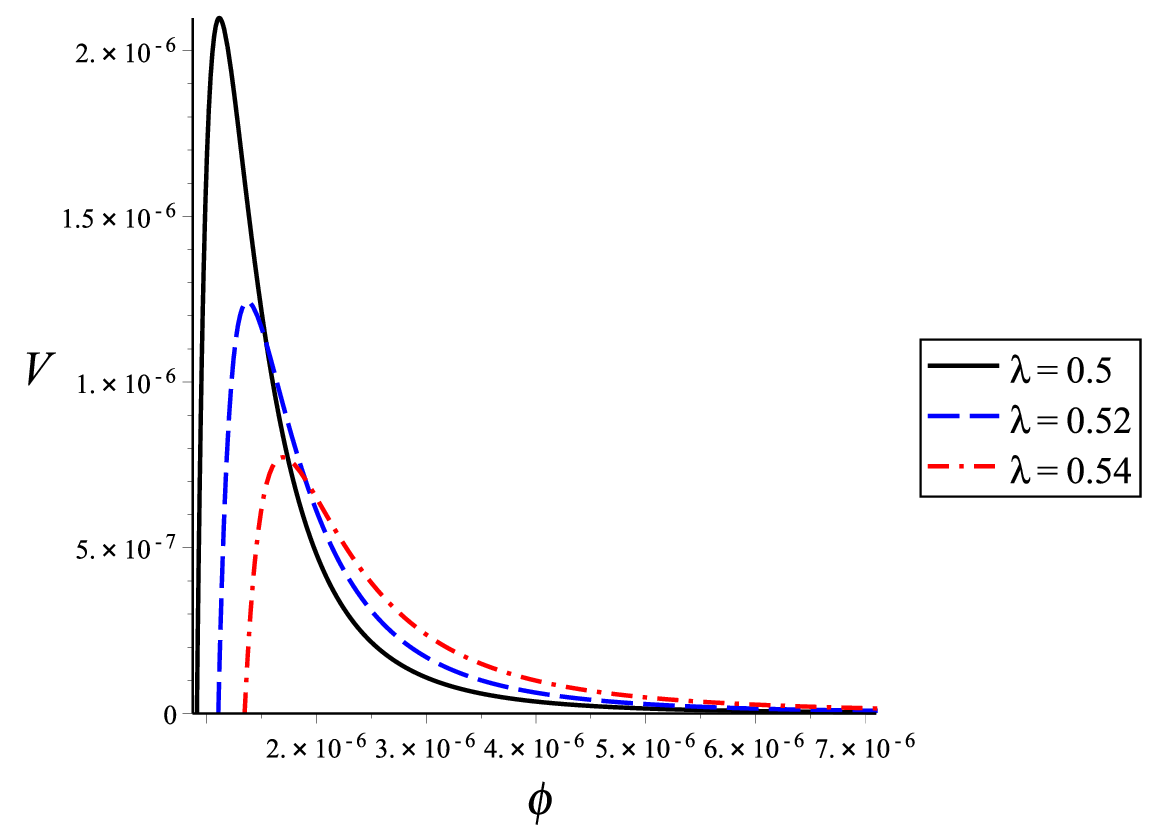}}
    \end{center}
   \caption{\footnotesize The inflaton potential $V(\varphi)=\textbf{\textit{V}}/M_p^4$ versus the scalar field $\phi={\varphi/M_p}$ for $M=10^{-7}M_p$ and different values of $\lambda$.}
\label{fig2}
\end{figure}

\begin{figure}[t]
\begin{center}
  \scalebox{0.5}{\includegraphics{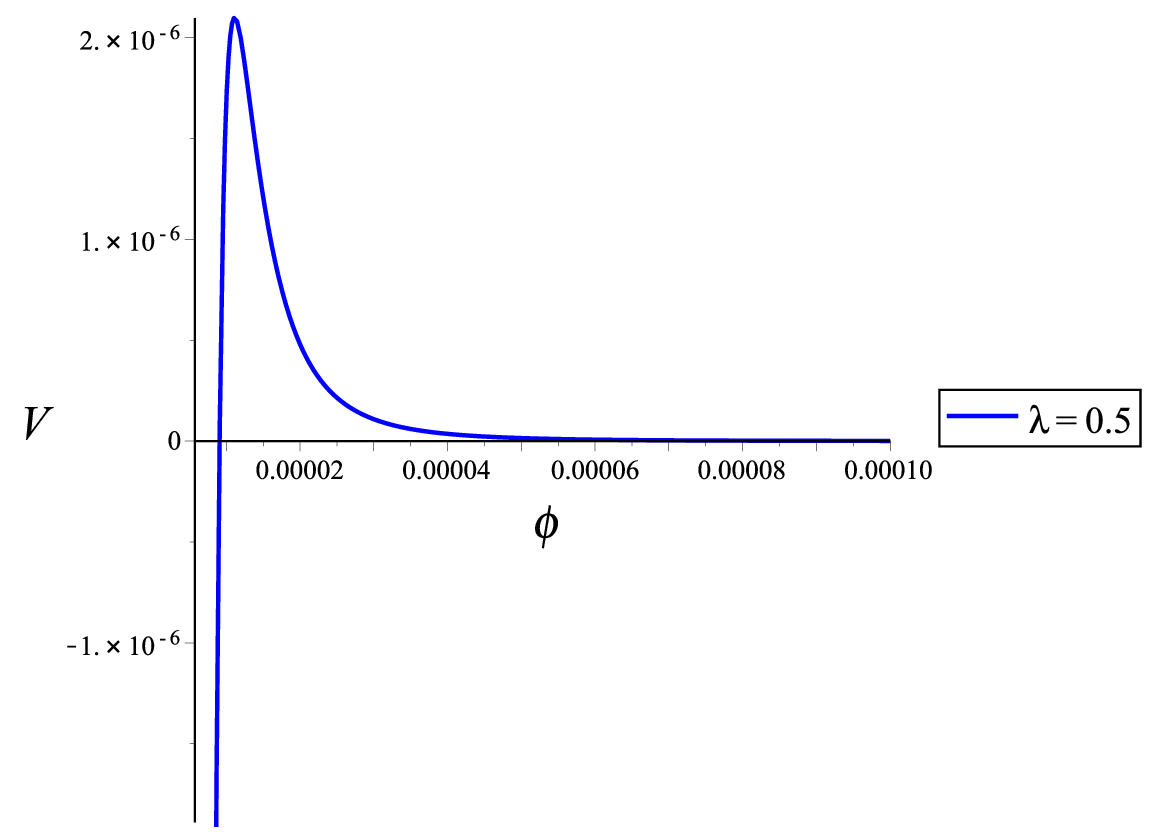}}
    \end{center}
  \caption{\footnotesize The inflaton potential $V(\varphi)=\textbf{\textit{V}}/M_p^4$ versus scalar field $\phi={\varphi/M_p}$ for $\lambda=0.5$ and $M=10^{-7}M_p$.}
\label{fig3}
\end{figure}

In this section we will investigate cosmological perturbation in GNMDC model.
Here, we will examine both scalar and tensor fluctuations during intermediate inflation.
To achieve this goal, we separate the space-time that represents our universe into two components: the background and the perturbation components.
The background is described by a homogeneous and isotropic FLRW metric, while the perturbed sector of the metric determines the anisotropic parts of the CMB.
To investigate the evaluation of cosmological perturbations in the GNMDC model, we need to derive the second-order action for the curvature perturbation $\mathcal{R}$ \cite{Dalianis,Chengjie,Germani2}
\begin{equation}\label{35}
\mathcal{S}_{(2)}={M_p^2\over2}\int d^4xa^3Q_s\Big[\dot{\mathcal{R}}^2-{c_s^2\over a^2}(\partial_i\mathcal{R})^2\Big],
\end{equation}
where $Q_s$ is defined as
\begin{equation}\label{36}
Q_s={M_p^2F^2\theta\over 3f(\varphi)H^2}\Big[1+3f(\varphi)H^2\Big({1+\theta\over1-\theta/3}\Big)\Big].
\end{equation}
Also, in this relation $F$ and $\theta$ are defined as
\begin{equation}\label{37}
 F={1-\theta/3\over 1-\theta},\;\;\; \theta\equiv{3\over2}{f(\varphi){\dot{\varphi}}^2\over M_p^2}.
\end{equation}

In relation (\ref{35}), ${c_s}^2$ is the sound speed squared for scalar modes, given by
\begin{equation}\label{38}
c_s^2={3\Big[1+\theta+3f(\varphi)H^2\Big(1+\theta+{4\theta\over9F}\Big)
+6\dot{H}f(\varphi)\big(1-\theta/3\big)\Big]\over\Big(3-\theta+9f(\varphi)H^2(1+\theta)\Big)}.
\end{equation}

At the time when the comoving wave number $k$ exits the horizon $c_sk=aH$, the power spectrum of the comoving curvature perturbation takes the following form

\begin{equation}\label{39}
\mathcal{P}_\mathcal{R}={H^2\over8\pi^2Q_sc_s^3}\Big|_{c_sk=aH}.
\end{equation}

As mentioned previously, by applying the high friction condition $f(\varphi)H^2\gg1$, the square of sound speed for scalar modes becomes

\begin{equation}\label{39.1}
c_s^2\approx{-7\theta^2+10\theta+9\over3(1+\theta)(3-\theta)}-{2(3-\theta)\over3(1+\theta)}\epsilon.
\end{equation}

Additionally, by using the slow roll condition $\epsilon\ll1$ we can rewrite the sound speed squared (\ref{38}) in the simple form of

\begin{equation}\label{40}
c_s^2\approx{-7\theta^2+10\theta+9\over3(1+\theta)(3-\theta)}.
\end{equation}

We have seen, that in the limit $\theta\rightarrow 0$, hence ${c_s}^2\approx1$. In the high friction regime the power spectrum (\ref{39}) becomes
\begin{equation}\label{41}
\mathcal{P}_\mathcal{R}={9\sqrt{3}(1-\theta)^2(1+\theta)^{(1/2)}(3-\theta)^{(1/2)}\over 8\pi^2\theta(-7\theta^2+10\theta+9)^{(3/2)}}\Big({H^2\over M_p^2}\Big).
\end{equation}
By differentiating the logarithm of the power spectrum with respect to the logarithm of $k$ at the horizon crossing $c_{s}k=aH$, one can obtain the spectral index as

\begin{equation}\label{42}
n_s-1={d\ln(\mathcal{P}_\mathcal{R})\over d\ln(k)}\Big|_{k=aH}.
\end{equation}

At the horizon crossing $d\ln{k}=H(1-\epsilon)dt$, thus the spectral index $n_s$ in the high friction condition becomes

\begin{eqnarray}\label{43}
&&n_s-1=\\ \nonumber
&&-\Big(3-{1\over2}{\theta\over(1+\theta)}+{1\over2}{\theta\over(3-\theta)}+2{\theta\over(1-\theta)}+{3\over2}
{(-14\theta+10)\theta\over(-7\theta^2+10\theta+9)}\Big)\epsilon \\ \nonumber
&&-\Big(1-{1\over2}{\theta\over(1+\theta)}+{1\over2}{\theta\over(3-\theta)}+2{\theta\over(1-\theta)}+{3\over2}
{(-14\theta+10)\theta\over(-7\theta^2+10\theta+9)}\Big)\delta_\mathcal{A}.
\end{eqnarray}

Another remarkable point, is the investigation of tensor perturbations during intermediate inflation, which would generate gravitational waves.
The tensor power spectrum can be written as \cite{Dalianis}

\begin{equation}\label{44}
\mathcal{P}_t={H^2\over2\pi^2Q_t c_t^3}\Big|_{k=aH},
\end{equation}

where $Q_t$ and the sound speed squared for tensor modes ${c_s}^2$ are given by the expressions
\begin{equation}\label{45}
Q_t={M_p^2\over4}\Big(1-{\theta\over3}\Big),\;\;\;\; c_t^2\approx1+{2\theta\over3}.
\end{equation}

Therefore in the high friction regime the power spectrum for tensor perturbations becomes
\begin{equation}\label{46}
\mathcal{P}_t\approx{18\sqrt{3} H^2\over\pi^2M_p^2(3-\theta)(3+2\theta)^{(3/2)}}.
\end{equation}
Hence, the tensor to scalar ratio $r$ in this regime, becomes
\begin{equation}\label{47}
r={\mathcal{P}_t\over\mathcal{P}_\mathcal{R}}\approx{16\theta(-7\theta^2+10\theta+9)^{(3/2)}\over (1-\theta)^2 (1+\theta)^{(1/2)}(3-\theta)^{(3/2)}(3+2\theta)^{(3/2)}}.
\end{equation}

\begin{figure}[t]
\begin{center}
  \scalebox{0.5}{\includegraphics{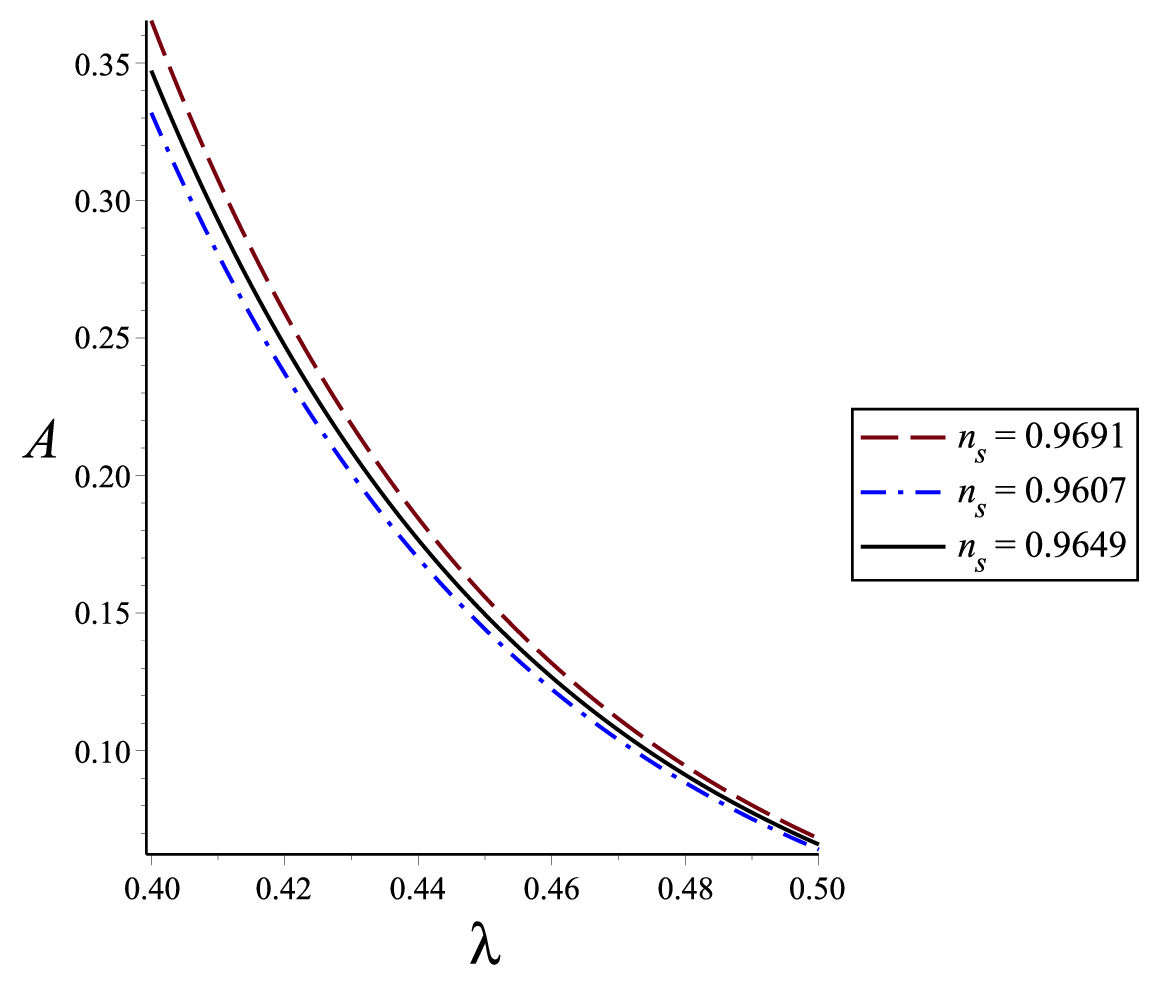}}
  \end{center}
   \caption{\footnotesize The evolution of parameter $A$ versus the $\lambda$, for different values of spectral index $n_s$. In this plot we assume that $M_p=1$.}
\label{fig4}
\end{figure}

\section{Numerical analysis}

\begin{figure}[t]
\begin{center}
  \scalebox{0.5}{\includegraphics{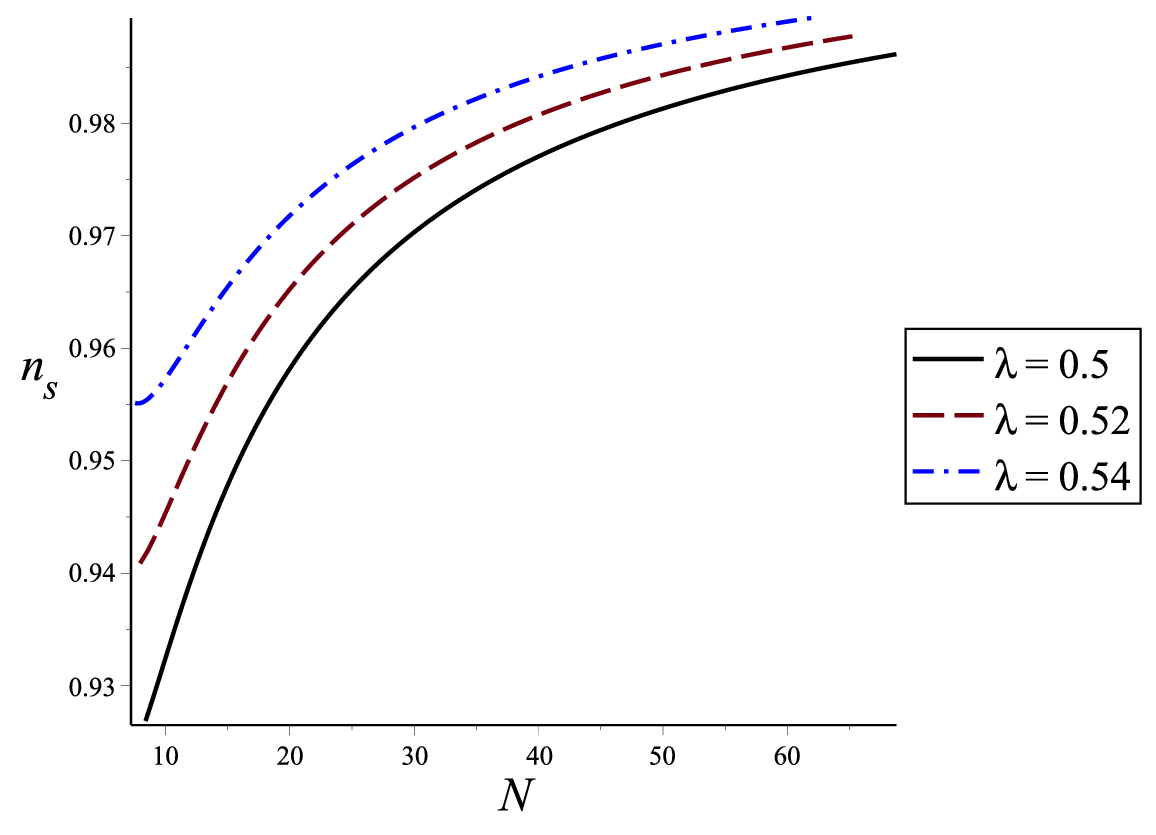}}
    \end{center}
   \caption{\footnotesize The evolution of scalar spectral index $n_s$ versus the number of e-folds, for different values of $\lambda$.
We assumes coupling constant $M=10^{-7}M_p$.}
\label{fig5}
\end{figure}

\begin{figure}[t]
\begin{center}
  \scalebox{0.5}{\includegraphics{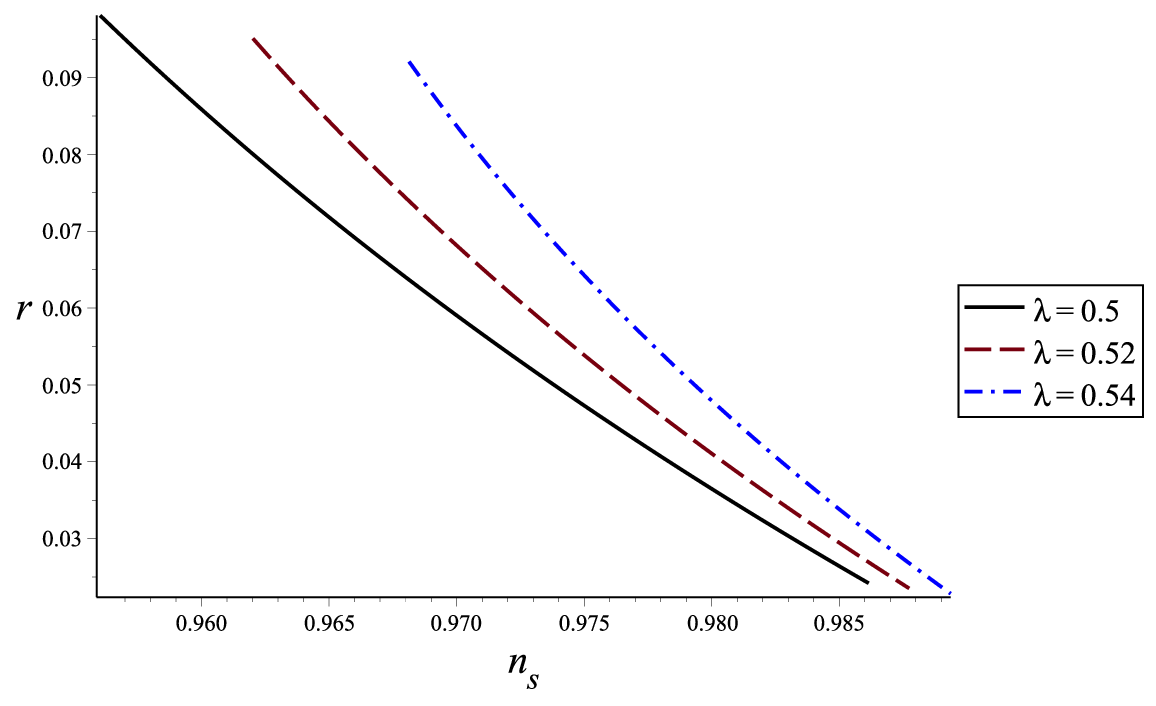}}
    \end{center}
   \caption{\footnotesize Tensor to scalar ratio $r$ versus the spectral index $n_s$ for different values of $\lambda$.}
\label{fig6}
\end{figure}

In the previous sections, we explored the intermediate inflation in the generalized non-minimal derivative coupling model with a power law coupling function.
 From the analytical investigation we have seen, that there are six parameters in this model, $\lambda$ and $A$ in the intermediate scale factor (\ref{17}), $M$ and $\alpha$ in the coupling function (\ref{13}), $c$ and $m$ in the power law scalar field (\ref{19}).
 For any values of $\lambda$ we can obtain $m$ and $\alpha$ from relation (\ref{23}). Therefore, there are four parameters that can be determined using perturbation equations and cosmological data.
 Now, in this section, our aim is to evaluate these four parameters in a way that not only satisfies the background constraints, such as the slow roll approximations and the high friction limit, but also ensures that the perturbational quantities are consistent with the observational data obtained from Planck 2018 data \cite{Planck}.

Therefore we will derive the parameters of the model as a function of cosmological data and evaluate these parameters using Planck 2018 data.
Thus, from relations (\ref{13}), (\ref{18}), (\ref{22}) and (\ref{37}), we can obtain $\theta$ as a function of $t$ in the high friction limit

\begin{equation}\label{48}
\theta={(1-\lambda)\over\lambda A t^\lambda}.
\end{equation}

By comparing relations (\ref{27}), (\ref{28}) with equation (\ref{48}) in the high friction regime, we deduce that $\theta=\epsilon=3(1-\lambda)\delta_{\mathcal{A}}$.
Therefore, when $\epsilon$ is much smaller than 1 in the slow roll condition, it implies that $\theta$ is also much smaller than 1.
In the high friction regime we can rewrite relations (\ref{41}), (\ref{43}) and (\ref{47}) in the following form
\begin{eqnarray}
\label{49}
\mathcal{P}_\mathcal{R}&\approx&{H^2\over8\pi^2M_p^2\theta}\approx {(A\lambda)^3\over8\pi^2M_p^2(1-\lambda)}t_\star^{(3\lambda-2)}, \\
\label{50}
n_s-1&\approx& -3(\epsilon+\delta_{\mathcal{A}})\approx {3\lambda-2\over A\lambda t_\star^\lambda}, \\
\label{51}
r&\approx&16\theta\approx {16(1-\lambda)\over\lambda A t_\star^\lambda}.
\end{eqnarray}

Three relations above, evaluated at the horizon crossing, which we represent with $t_\star$. Now, by using relations (\ref{49}) and (\ref{50}), we can calculate the time of horizon crossing $t_\star$ as a function of the inflationary observable as
\begin{equation}\label{51.1}
t_\star=\Big(8\pi^2M_p^2(1-\lambda)\mathcal{P}_\mathcal{R}\Big)^{-{1\over2}}\Big({1-n_s\over2-3\lambda}\Big)^{3\over2}.
\end{equation}

By substituting of $t_\star$ in relation (\ref{50}) we can obtain $A$ as a function of $\lambda$, $n_s$ and $\mathcal{P}_\mathcal{R}$ as
\begin{equation}\label{52}
A={\big[8\pi^2M_p^2(1-\lambda)\mathcal{P}_\mathcal{R}\big]^{\lambda\over2}\over \lambda}\Big({2-3\lambda\over 1-n_s}\Big)^{\big({2-3\lambda\over2}\big)}.
\end{equation}

Also, it is important to note that, according to the Planck 2018 results \cite{Planck}, the value of $1-n_s$ is greater than $0$.
 Therefore, we require $0<\lambda<2/3$ based on relation (\ref{50}).
At the CMB scale $k_\star=0.05Mpc^{-1}$ the results of Planck 2018 \cite{Planck} provide the following constraints on the power spectrum, the scalar spectral index and the tensor to scalar ratio as
\begin{eqnarray}
\label{53}
\ln(10^{10}\mathcal{P}_\mathcal{R})&=&3.044\pm0.014 \;\;\;\; (68\%\; C.L.),\\
\label{54}
n_s&=&0.9649 \pm 0.0042\;\;\; (68\% \; C.L.), \\
\label{55}
r&<&0.07\;\;\;\;\;\; (95\%\; C.L.).
\end{eqnarray}

Now, using this data and relation (\ref{52}), we can evaluate $A=0.066M_p^{1/2}$ for $\lambda=0.5$.
From constraints (\ref{33.1}) and (\ref{33.2}), for $\lambda=0.5$ the values of parameters $M$ and $c$ must be constrained as
\begin{eqnarray}
\label{56}
M&\ll&0.1796 M_p,\\
\label{57}
c&\ll&0.4193{M_p}^{5/4}.
\end{eqnarray}

Now, to display the high friction condition $f(\varphi)H^2\gg1$ in our plots, we can parameterize relation (\ref{33}) as follows
\begin{equation}\label{57}
1\ll Z\equiv{(A\lambda)^2 c^{(\alpha-1)}\over M^{(\alpha+1)}}.
\end{equation}
In the following, by substituting $c$ from relation (\ref{33}) into equation (\ref{57}) we obtain
\begin{equation}\label{57.1}
Z=(A\lambda)^2 \Bigg({8M_p^2(1-\lambda)\over3A\lambda^3}\Bigg)^{2(1-\lambda)\over(2-\lambda)}M^{-2}.
\end{equation}

As you can see, the high friction condition restricts the range of values for the $M$ and $c$ parameters given specific values of $\lambda$.

In Figure \ref{fig1}, we display the admissible regions for the values of $c$ and $M$ parameters, under the high friction constraint $(Z\gg1)$.
Various ranges of $Z$ parameters are represented by different shades of violet color.
The light-colored region indicates low values of $Z$, whereas the dark-colored region indicates high values of $Z$.
Also, the acceptable values for $c$ and $M$ parameters from relation (\ref{33}) are shown with a red dashed line.
The best values for parameters $c$ and $M$ are indicated by the red dashed line, which is located within the shaded region.

Thus, we chose $M=10^{-7}M_p$ from the dark violet region of Figure \ref{fig1} for $\lambda=0.5$.
Now, by using equation (\ref{33}), we can calculate the value of $c$ as $c =2.334\times10^{-7}M_p^{(5/4)}$.
Furthermore, using equation (\ref{57}), we determine a high friction parameter of $Z=3.226\times10^{12}$, consistent with the high friction regime where $Z\gg1$.

In Figure \ref{fig2} we plot the inflationary potential versus the scalar field for $M=10^{-7}M_p$ and different values of the parameter $\lambda$ from the beginning of inflation until the end of inflation.
As evident, the potential reaches its maximum value at $\varphi_{max}$, where the magnitude of the maximum potential decreases as the $\lambda$ parameter increases.

\begin{figure}[t]
\begin{center}
  \scalebox{0.5}{\includegraphics{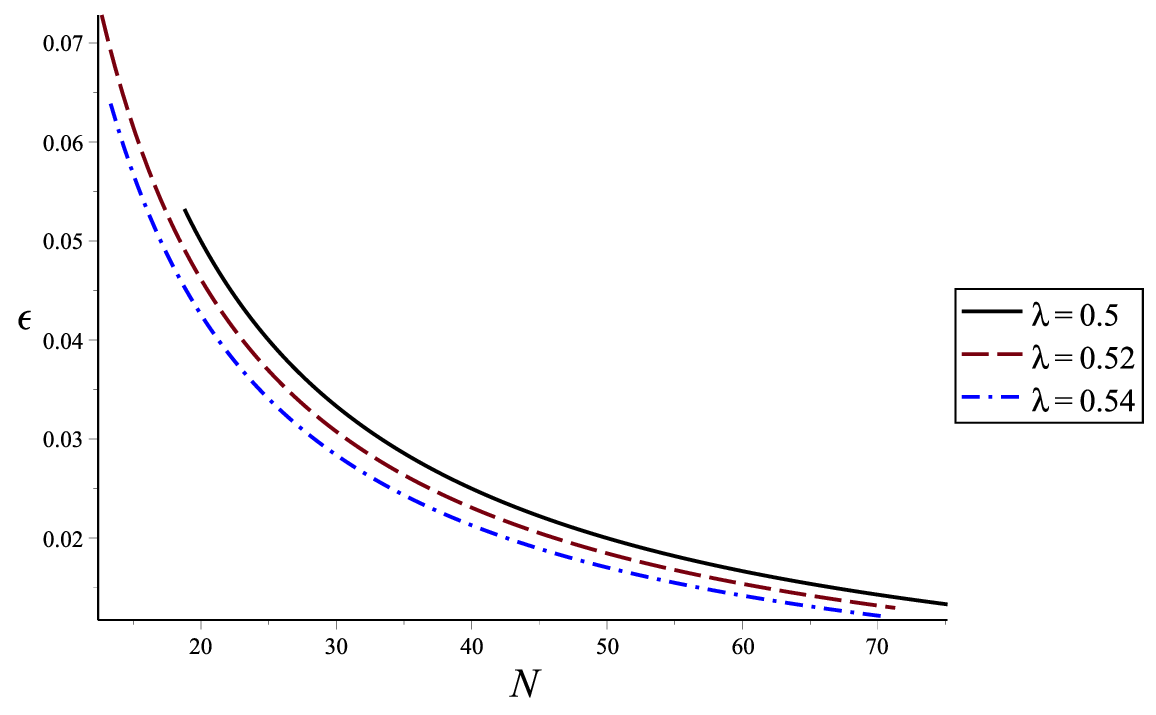}}
    \end{center}
   \caption{\footnotesize The slow roll parameter $\epsilon$ versus the number of e-folds $N$ for different values of $\lambda$.}
\label{fig7}
\end{figure}

After selecting the model parameters, such as $\lambda=0.5$ and $M=10^{-7}M_p$, the potential can be expressed as a function of the scalar field as
\begin{equation}\label{58}
V(\varphi)\approx-1.6\times10^{-42}{M_p^8\over \phi^6}\Big(0.5M_p^2-6.05\times10^9\phi^2\Big).
\end{equation}
We depicted the inflationary potential as a function of the scalar field for $M=10^{-7}M_p$ and $\lambda=0.5$ in Figure \ref{fig3}, separately.
The time of beginning and the ending of inflation in this model are $t_0=2.30\times10^{2}M_p^{-1}$ and $t_{end}=8.562\times 10^5 M_p^{-1}$ respectively.
Therefore, the values of scalar field at the beginning and the ending of inflation are $\varphi_0=9.092\times 10^{-7}M_p$ and $\varphi_{end}=7.1\times 10^{-6}M_p$ respectively.
Additionally, the maximum value of the potential occurs at the $\varphi_{max}=1.11\times 10^{-6}M_p$ where the potential reaches its maximum value
$V(\varphi_{max})=2.098\times10^{-6}M_p^4$.
In the same way, we can obtain the potential at the beginning and the end of inflation as $V(\varphi_0)=0$ and $V(\varphi_{end})=3.74\times10^{-9}M_p^4$ respectively.
Subsequently, the time of horizon crossing can be obtained from relation (\ref{51.1}) as $t_\star=1.868\times10^{5}M_p^{-1}$. The scalar field value at horizon crossing is $\varphi_{\star}=4.85\times10^{-6}M_p$ and the potential at that time is $V(\varphi_{\star})=1.68\times10^{-8}M_p^4$ for $\lambda=0.5$.

In Table 1, we have gathered the parameters of the model, to compare the values of different quantities, consistent with observational data for $\lambda=0.4$, $\lambda=0.5$, and $\lambda=0.6$.

\begin{center}
\begin{tabular}[t]{|c|c|c|c|}
\hline
 \multicolumn{4}{ |c| }{Table 1: Evaluation of the model's parameters} \\
\hline
 &$\lambda=0.4$ & $\lambda=0.5$ & $\lambda=0.6$ \\
\hline
$A$ & $0.347{M_P}^{2/5}$ & $0.066{M_P}^{1/2}$ & $0.014{M_P}^{3/5}$ \\
\hline
$\alpha$ & $7$ & $5$ & $3.66$ \\
\hline
$M$ & $10^{-7} M_p$ & $10^{-7} M_p$ & $10^{-7} M_p$ \\
\hline
$10^7c$ &$1.70{M_p}^{6/5}$ & $2.33{M_p}^{5/4}$ & $3.52{M_p}^{13/10}$ \\
\hline
$Z$ & $4.76\times10^{13}$ & $3.226\times10^{12}$ & $2.00\times10^{11}$ \\
\hline
$M_p t_\star$ & $3.45\times10^5$ & $1.87\times10^5$ & $0.53\times10^5$ \\
\hline
$10^{7}M_p^{-1}\varphi_0$ & $3.55$ & $9.09$ & $0.24$ \\
\hline
$10^{7}M_p^{-1}\varphi_{end}$ & $22.71$ & $71.01$ & $232.21$ \\
\hline
$M_pt_{0}$ & $38.79$ & $230.1$ & $631.0$ \\
\hline
$10^{-5}M_pt_{end}$ & $4.18$ & $8.56$ & $11.62$ \\
\hline
\end{tabular}
\end{center}

In Figure \ref{fig4} we have illustrated the evolution of $A$ versus the $\lambda$ for different values of the spectral index $n_s$.
 We have seen, in the range $0<\lambda<2/3$, the value of $A$ is within the interval $A\in(0,\infty)$.

Figure \ref{fig5} illustrates the evolution of the scalar spectral index $n_s$ versus the number of e-folds, for different values of the parameter $\lambda$ that corresponds to the intermediate expansion of the scale factor.

In Figure \ref{fig6}, the tensor to scalar ratio $r$ is plotted against the spectral index $n_s$ for various values of $\lambda$.

As we can see, with the appropriate choice of the model's parameters, Intermediate inflation in GNMDC is consistent with observational results of Planck 2018 \cite{Planck}.

Moreover, we observed that for different values of $\lambda$, the predictions of our model are well corroborated by the region $95\%CL$ Planck 2018 data.

In Figure \ref{fig7} the slow roll parameter $\epsilon$ versus the number of e-folds $N$ for different values of $\lambda$ has been depicted.
This figure clearly shows that the slow roll parameter $\epsilon$ starts at 1 as the beginning of inflation,
and decreases throughout inflation until the end of inflation, in contrast to the standard canonical inflation.

Therefore, the problem of a graceful exit cannot be resolved simultaneously with the end of inflation. Consequently, we must make modifications to the potential after inflation period, in order to address this issue.

According to the observational data the energy density scale at the horizon exit is approximately $V_{\star}^{1/4}\approx10^{-2}M_p$. This value is two orders of magnitude lower than the Planck energy scale which is in good agreement with the energy scale of big bang cosmology.

\section{conclusion}

In this work, we have considered an intermediate inflationary model in the context of the generalized non-minimal derivative coupling with the power law coupling function.
We have obtained power law solutions for Friedmann's equations of GNMDC, in the slow roll regime using the exponential function scale factor.
Also, we have identified all of the constraints on the parameters of the model in the background equations based on the high friction limit and slow roll approximation.
Moreover, by investigating the evaluation of scalar and tensor perturbations during intermediate inflation, we calculate the scalar power spectrum, tensor power spectrum, scalar spectral index and tensor to scalar ratio for intermediate inflation in GNMDC model.
In this sense, we have evaluated the parameters of the model by comparing the perturbational quantity from our model with the Planck 2018 results.
We demonstrate that by selecting the appropriate parameters for the model we can generate the observational data such as the spectral index, power spectrum, and tensor-to-scalar ratio that are consistent with the Planck 2018 data.
We find the admissible regions for the values of the parameters $c$ and $M$ in a way that not only satisfies the background constraints, such as the slow roll approximations and the high friction limit, but also ensures that the perturbational quantities are consistent with the observational data.
We have computed the suitable effective potential, with zero defined at the onset of inflation, leading to intermediate inflation in GNMDC model.
The intermediate inflation in the GNMDC model is a successful theory that accurately describes a range of physical quantities associated with inflation, including the inflaton potential, energy scale, horizon crossing time, and number of e-folds. This model not only satisfies background constraints but also agrees with the results of Planck 2018.


\begin{thebibliography}{99}

\bibitem{Mukhanov} Viatcheslav F. Mukhanov, H.A. Feldman, Robert H. Brandenberger, Theory of cosmological perturbations, Phys.Rept. 215 (1992) 203.
\bibitem{Olive} K.A. Olive, Inflation, Phys. Rept. 190 (1990) 307.
\bibitem{guth} A. H. Guth, Inflationary universe: A possible solution to the horizon and flatness problems, Phys. Rev. D 23, 347 (1981).
\bibitem{Linde1} A.D. Linde, A New Inflationary Universe Scenario: A Possible Solution of the Horizon, Flatness, Homogeneity, Isotropy and Primordial Monopole Problems, Phys. Lett. B 108 (1982) 389.
\bibitem{Linde2} ] A.D. Linde, Chaotic Inflation, Phys. Lett. B 129 (1983) 177.
\bibitem{Linde3} A. D. Linde, Particle physics and inflationary cosmology, Vol. 5. 1990. arXiv: hep-th/0503203.
\bibitem{kolb} E. W. Kolb and M. S. Turner. The Early Universe. 90. isbn: 978-0-201-62674-2. doi: 10.1201/9780429492860.
\bibitem{Lyth} D.H. Lyth and A. Riotto, Particle physics models of inflation and the cosmological density perturbation, Phys. Rept. 314 (1999) 1.
\bibitem{Liddle} David Wands, Karim A. Malik, David H. Lyth, Andrew R. Liddle, A new approach to the evolution of cosmological perturbations on large scales, Phys. Rev. D 62 (2000) 043527.
\bibitem{Liddle1} A.R. Liddle and D.H. Lyth, Cosmological inflation and large scale structure, Cambridge University Press (2000), 10.1017/CBO9781139175180.
\bibitem{Odintsov1} S. D. Odintsov, V. K. Oikonomou, I. Giannakoudi, F. P. Fronimos and E. C. Lymperiadou, Recent Advances in Inflation, Symmetry 15 (2023), 1701.
\bibitem{Martin} J. Martin, C. Ringeval and V. Vennin, Encyclopædia Inflationaris, Phys. Dark Univ. 5-6 (2014) 75–235.
\bibitem{Bardeen} J.M. Bardeen, P.J. Steinhardt and M.S. Turner, Spontaneous Creation of Almost Scale - Free Density Perturbations in
an Inflationary Universe, Phys. Rev. D 28 (1983) 679.
\bibitem{Odintsov2} S. Nojiri, S. D. Odintsov and V. K. Oikonomou, Modified Gravity Theories on a Nutshell: Inflation, Bounce and Late-time Evolution, Phys. Rept. 692 (2017), 1.
\bibitem{Odintsov3} S. Nojiri and S. D. Odintsov, Unified cosmic history in modified gravity: from F(R) theory to Lorentz non-invariant models, Phys. Rept. 505 (2011), 59.
\bibitem{Mukhanov1} T. Damour and V. F. Mukhanov, Inflation without Slow Roll, Phys. Rev. Lett. 80, 3440, (1998).
\bibitem{Capozziello} Salvatore Capozziello, Mariafelicia De Laurentis, Extended Theories of Gravity, Phys. Rept. 509 (2011) 167.
\bibitem{Peebles} Bharat Ratra, P.J.E. Peebles, Cosmological Consequences of a Rolling Homogeneous Scalar Field, Phys. Rev. D 37 (1988) 3406.
\bibitem{Stephani} H. Stephani, D. Kramer, M. A. H. MacCallum, C. Hoenselaers and E. Herlt, Exact solutions of Einstein’s field equations, Cambridge Univ. Press, 2003, ISBN 978-0-521-46702-5.
\bibitem{Lucchin} F. Lucchin, S. Matarrese, Power-law inflation, Phys. Rev. D 32, 1316 (1985).
\bibitem{Barrow} John D. Barrow, Phys. Lett. B 235, 40 (1990).
\bibitem{Barrow1} John D. Barrow, Andrew R. Liddle, and Cédric Pahud, Intermediate inflation in light of the three-year WMAP observations, Phys. Rev. D 74, 127305 (2006).
\bibitem{Herrera} Carlos González, Ramón Herrera, Intermediate inflation in a generalized induced-gravity scenario, Eur. Phys. J. C 77, 648(2017).
\bibitem{Herrera1} S. del Campo, R. Herrera, Intermediate inflation on the brane, Phys. Lett. B 670, 266 (2009).
\bibitem{Herrera2} Ramón Herrera, Nelson Videla, Marco Olivares, G-inflation: from the intermediate, logamediate and exponential models, Eur. Phys. J. C 78, 934 (2018).
\bibitem{Herrera3} Ramón Herrera, Nelson Videla, Marco Olivares, Warm intermediate inflationary Universe model in the presence of a generalized Chaplygin gas, Eur. Phys. J. C 76, 35 (2016).
\bibitem{Herrera4} R. Herrera, N. Videla and M. Olivares, Warm G inflation: Intermediate model, Phys. Rev. D 100, 023529 (2019).
\bibitem{Herrera5} R. Herrera, E. San Martin, Warm-intermediate inflationary universe model in braneworld cosmologies, Eur. Phys. J. C 71, 1701 (2011).
\bibitem{Rezazadeh} K. Rezazadeh, K. Karami, P. Karimi, Intermediate inflation from a non-canonical scalar field, JCAP 09 (2015) 053.
\bibitem{Teimoori} Zeinab Teimoori, and Kayoomars Karami, Galileon Intermediate Inflation, The Astrophysical Journal, 864, 41, (2018).
\bibitem{Rashidi} Narges Rashidi, Intermediate and Power-law Inflation in the Tachyon Model with Constant Sound Speed, Astrophys. J. 933, 46, (2022).
\bibitem{Horndeski} G. W. Horndeski, Second-order scalar-tensor field equations in a four-dimensional space, Int. J. Theor. Phys. 10, 363 (1974).
\bibitem{Charmousis} C. Charmousis, E. J. Copeland, A Padilla and P. M. Saffin, General Second-Order Scalar-Tensor Theory and Self-Tuning, Phys. Rev. Lett. 108, 051101 (2012).
\bibitem{Nakayama} K. Nakayama, F. Takahashi, Running kinetic inflation, JCAP, 11, 009 (2010).
\bibitem{Kobayashi} Tsutomu Kobayashi, Masahide Yamaguchi, Jun'ichi Yokoyama, Generalized G-inflation: Inflation with the most general second-order field equations, Prog. Theor. Phys. 126 (2011) 511-529.
\bibitem{Deffayet} C. Deffayet, S. Deser, G. Esposito-Farese, Generalized Galileons: All scalar models whose curved background extensions maintain second-order field equations and stress-tensors, Phys. Rev. D 80 (2009) 064015.
\bibitem{Deffayet1} C. Deffayet, Gilles Esposito-Farese, A. Vikman, Covariant Galileon, Phys. Rev. D 79 (2009) 084003.
\bibitem{Brans} C. Brans, R. H. Dicke, Mach's Principle and a Relativistic Theory of Gravitation, Phys. Rev. 124, 925 (1961).
\bibitem{Sushkov1} A. A. Starobinsky, S. V. Sushkov and M. S. Volkov, The screening Horndeski cosmologies, JCAP 06 (2016), 007.
\bibitem{Sushkov2} A. A. Starobinsky, S. V. Sushkov and M. S. Volkov, Anisotropy screening in Horndeski cosmologies, Phys. Rev. D 101, 064039 (2020).
\bibitem{Germani1} C. Germani and A. Kehagias, New Model of Inflation with Nonminimal Derivative Coupling of Standard Model Higgs Boson to Gravity, Phys. Rev. Lett. 105, 011302.
\bibitem{Germani2} C. Germani and A. Kehagias, Cosmological perturbations in the new Higgs inflation, JCAP 05(2010)019.
\bibitem{Germani3} C. Germani and Y. Watanabe, UV-protected (natural) inflation: primordial fluctuations and non-gaussian features, JCAP 07(2011)031.
\bibitem{Tsujikawa} Shinji Tsujikawa, Observational tests of inflation with a field derivative coupling to gravity, Phys. Rev. D, 85, 083518 (2012).
\bibitem{Sadjadi} H. Mohseni Sadjadi, P. Goodarzi, Reheating in nonminimal derivative coupling model, JCAP 1302 (2013) 038.
\bibitem{Sadjadi1} H. Mohseni Sadjadi, P. Goodarzi, Temperature in warm inflation in non minimal kinetic coupling model, Eur. Phys. J. C 75 (2015) 513.
\bibitem{Parviz} P. Goodarzi, Gravitational baryogenesis in non-minimal kinetic coupling model, Eur. Phys. J. C 83 (2023) 990.
\bibitem{Sushkov3} S. V. Sushkov, Realistic cosmological scenario with nonminimal kinetic coupling, Phys. Rev. D 85, 123520 (2012).
\bibitem{Sushkov4} S. V. Sushkov, Exact cosmological solutions with nonminimal derivative coupling, Phys. Rev. D 80, 103505 (2009).
\bibitem{Sushkov5} E. N. Saridakis and S. V. Sushkov, Quintessence and phantom cosmology with nonminimal derivative coupling, Phys. Rev. D 81, 083510 (2010).
\bibitem{Sushkov6} S. V. Sushkov and R. Galeev, Cosmological models with arbitrary spatial curvature in the theory of gravity with non-minimal derivative coupling, Phys. Rev. D 108, 044028 (2023).
\bibitem{Chengjie} Chengjie Fu, Puxun Wu, Hongwei Yu, Primordial Black Holes from Inflation with Nonminimal Derivative Coupling, Phys. Rev. D 100 (2019) 063532.
\bibitem{Dalianis} Ioannis Dalianis, Stelios Karydas, Eleftherios Papantonopoulos, Generalized Non-Minimal Derivative Coupling: Application to Inflation and Primordial Black Hole Production, JCAP 06 (2020) 040.
\bibitem{Tiwari} Yashi Tiwari, Nilanjandev Bhaumik, Rajeev Kumar Jain, Understanding large scale CMB anomalies with the generalized nonminimal derivative coupling during inflation, Phys. Rev. D 107, 103513.
\bibitem{Planck} Y. Akrami et al. (Planck Collaboration), Planck 2018 results. X. Constraints on inflation, Astron. Astrophys. 641 (2020) A10.
\bibitem{Martin1} J. Martin and C.Ringeval, First CMB constraints on the inflationary reheating temperature, Phys. Rev. D 82, 023511 (2010).
\bibitem{Gialamas} I. D. Gialamas, A. Karam, A. Lykkas and T. Pappas, Palatini-Higgs inflation with nonminimal derivative coupling, Phys. Rev. D 102 (2020) 063522.
\end{thebibliography}
\end{document}